\documentclass[aps,prc,twocolumn,superscriptaddress,groupedaddress]{revtex4}  
\usepackage{graphicx}  
\usepackage{dcolumn}   
\usepackage{bm}        

\usepackage{threeparttable} 
\usepackage{caption}

\usepackage[cmex10]{amsmath}

\usepackage{url}
\usepackage{natbib}

\usepackage{siunitx}
\usepackage{enumerate}
\usepackage{subfigure}
\usepackage{tabularx}
\usepackage{listings}
\usepackage{color}
\definecolor{mygreen}{RGB}{28,172,0} 
\definecolor{mylilas}{RGB}{170,55,241}
\usepackage{xcolor}
\makeatletter
\AtBeginDocument{%
  \let\c@figure\c@lstlisting
  
  \let\ftype@lstlisting\ftype@figure 
}
\makeatother

\lstdefinestyle{matlabFile}{
    language=Matlab,
    tabsize=3,
    caption=Test,
    label=code:sample,
    frame=shadowbox,
    rulesepcolor=\color{gray},
    xleftmargin=20pt,
    framexleftmargin=15pt,
    commentstyle=\color{mygreen}, 
    stringstyle=\color{black}, 
    numbers=left,
    numberstyle=\tiny,
    numbersep=5pt,
    breaklines=true,
    showstringspaces=false,
    basicstyle=\ttfamily\footnotesize\bfseries, 
    basewidth  = {.5em,0.4em},
    otherkeywords={set, run, foreach, let, return}, 
    keywordstyle=\color{blue}\bf,
    emph={Test,Message,Implementation, Process_Rank, Num_Processes, avg, standard_deviation, median, max, min, shmem_get, mpi_get, shmem_put, mpi_put, num_pes, rank, mpi_sendrecv, mpi_send_recv, mpi_isend_irecv, Root, HPC_EXP, tr, td, table},emphstyle={\color{magenta}}}

\begin{document}

\title{Deep learning: Extrapolation tool for \textit{ab initio} nuclear theory} 

\author{Gianina Alina Negoita} 
\affiliation{Department of Computer Science, Iowa State University, Ames, Iowa  50011, USA}
\affiliation{Horia Hulubei National Institute for Physics and Nuclear Engineering, Bucharest-Magurele 76900, Romania}
\author{James P. Vary} 
\affiliation{Department of Physics and Astronomy, Iowa State University, Ames, Iowa  50011, USA}
\author{Glenn R. Luecke} 
\affiliation{Department of Mathematics, Iowa State University, Ames, Iowa 50011, USA}
\author{Pieter Maris} 
\affiliation{Department of Physics and Astronomy, Iowa State University, Ames, Iowa 50011, USA}
\author{Andrey M. Shirokov} 
\affiliation{Skobeltsyn Institute of Nuclear Physics, Moscow State University, Moscow 119991, Russia}
\affiliation{Department of Physics, Pacific National University, Khabarovsk 680035, Russia}
\author{Ik Jae Shin} 
\affiliation{Rare Isotope Science Project, Institute for Basic Science, Daejeon 34047, Korea}
\author{Youngman Kim} 
\affiliation{Rare Isotope Science Project, Institute for Basic Science, Daejeon 34047, Korea}
\author{Esmond G. Ng}
\affiliation{Lawrence Berkeley National Laboratory, Berkeley, California 94720, USA}
\author{Chao Yang}
\affiliation{Lawrence Berkeley National Laboratory, Berkeley, California 94720, USA}
\author{Matthew Lockner} 
\affiliation{Department of Physics and Astronomy, Iowa State University, Ames, Iowa 50011, USA}
\author{Gurpur M. Prabhu} 
\affiliation{Department of Computer Science, Iowa State University, Ames, Iowa  50011, USA}

\date{\today}

\begin{abstract} 
\textit{Ab initio} approaches in nuclear theory, such as the no-core shell model (NCSM), have been developed for approximately solving finite nuclei with realistic strong interactions. The NCSM  and other approaches require an extrapolation of the results obtained in a finite basis space to the infinite basis space limit and assessment of the uncertainty of those extrapolations.  Each observable requires a separate extrapolation and many observables have no proven extrapolation method. We propose a feed-forward artificial neural network (ANN) method as an extrapolation tool to obtain the ground-state energy and the ground-state point-proton root-mean-square (rms) radius along with their extrapolation uncertainties. The designed ANNs are sufficient to produce results for these two very different observables in $^6\mathrm{Li}$ from the \textit{ab initio} NCSM results in small basis spaces that satisfy the following theoretical physics condition: independence of basis space parameters in the limit of extremely large matrices. Comparisons of the ANN results with other extrapolation methods are also provided.
\end{abstract}


\maketitle

\section{Introduction} \label{sec:Introduction}
A major long-term goal of nuclear theory is to understand how low-energy nuclear properties arise from strongly interacting nucleons.  When interactions that describe nucleon-nucleon (NN) scattering data with high accuracy are employed, the approach is considered to be a first principles or \textit{ab initio} method.  This challenging quantum many-body problem requires a non-perturbative computational approach for quantitative predictions.

With access to powerful high-performance computing (HPC) systems, several \textit{ab initio} approaches have been developed to study nuclear structure and reactions. The no-core shell model (NCSM)~\cite{NCSM:2013} is one of these approaches that falls into the class of configuration interaction methods. \textit{Ab initio} theories, such as the NCSM, traditionally employ realistic inter-nucleon interactions 
and provide predictions for binding energies, spectra, and other observables in light nuclei. 

The NCSM casts the non-relativistic quantum many-body problem as a finite Hamiltonian matrix eigenvalue problem expressed in a chosen, but truncated, basis space. A popular choice of basis representation is the three-dimensional harmonic-oscillator (HO) basis that we employ here. This basis is characterized by the HO energy, $\hbar\Omega$, and the many-body basis space cutoff, $N_{\rm max}$.
The $N_{\rm max}$ cutoff for the configurations to be included in the basis space is defined as the maximum of the sum over all nucleons of their HO quanta (twice the radial quantum number plus the orbital quantum number) above the minimum needed to satisfy the Pauli principle. Due to the strong short-range correlations of nucleons in a nucleus, a large basis space (model space) is required to achieve convergence in this two-dimensional parameter space ($\hbar\Omega$, $N_{\rm max}$), where convergence is defined as independence of both parameters within evaluated uncertainties. However, one faces major challenges to approach convergence since, as the size of the space increases, the demands on computational resources grow rapidly. In practice these calculations are limited and one can not directly calculate, for example, the ground-state (GS) energy or the GS point-proton root-mean-square (rms) radius for a sufficiently large $N_{\rm max}$ that would provide good approximations to the converged result in most nuclei of interest~\cite{Vary:2009qp, ExtrapolationB1:2009, ExtrapolationB2:2013, SHIROKOV:LightNuclei:JISP16:2014}.  We focus on these two observables in the current investigation.

To obtain the GS energy and the GS point-proton rms radius as close as possible to the exact results, the NCSM and other \textit{ab initio} approaches require an extrapolation of the results obtained in a finite basis space to the infinite basis space limit and assessment of the uncertainty of those extrapolations~\cite{ExtrapolationB1:2009, ExtrapolationB2:2013, ExtrapolationA:2017}. 
Each observable requires a separate extrapolation and many observables have no proposed extrapolation method at the present time.

Deep learning is a subfield of machine learning concerned with algorithms inspired by the structure and function of the brain called artificial neural networks (ANNs). In recent years, deep learning became a tool for solving challenging data analysis problems in a number of domains. For example, several successful applications of the ANNs have emerged in nuclear physics, high-energy physics, astrophysics, as well as in biology, chemistry, meteorology, geosciences, and other fields of science. Applications of ANNs to quantum many-body systems have involved multiple disciplines and have been under development for many years~\cite{Clark1999}. 

ANNs have been applied previously to an array of problems in nuclear physics.   For example, ANN models have been developed for identification of impact parameter in heavy ion collisions~\cite{DAVID:1995, BASS:1996, HADDAD:1997}, statistical modeling of nuclear systematics~\cite{CLARK:SMONS:2001}, developing nuclear mass systematics~\cite{NuclearMassSystematics:ATHANASSOPOULOS:2004}, determining one- and two-proton separation energies~\cite{ProtonSeparationEnergies:ATHANASSOPOULOS:2004}, modeling systematics of $\beta$ decay half-lives~\cite{COSTIRIS:betaDecayHalf-lives:2007, Costiris:2008pp}, constructing a model for the nuclear charge radii~\cite{AKKOYUN:2013}, and obtaining potential energy curves~\cite{AKKOYUN:potentialEnergyCurves:2013}.   More recent applications include predicting nuclear masses for properties of neutron stars~\cite{Utama:2015hva}, predicting nuclear charge radii~\cite{Utama:2016tcl}, as well as improving and validating nuclear mass formulas~\cite{Utama:2017wqe,Utama:2017ytc}.   An ambitious application of ANNs for extrapolating nuclear binding energies is also noteworthy~\cite{PhysRevC.98.034318}.

The present work proposes a feed-forward ANN method as an extrapolation tool to obtain the GS energy and the GS point-proton rms radius  and their extrapolation uncertainties based upon NCSM results in readily-solved basis spaces. The advantage of ANN is that it does not need an explicit analytical expression to model the variation of the GS energy or the GS point-proton rms radius with respect to $\hbar\Omega$ and $N_{\rm max}$. We will demonstrate that the feed-forward ANN method is very useful for estimating the converged result at very large $N_{\rm max}$ through demonstration applications in $^6\mathrm{Li}$.

We have generated theoretical data for $^6\mathrm{Li}$ by performing \textit{ab initio} NCSM calculations with the MFDn code~\cite{MFDN:STERNBERG:2008, MFDN:MARIS:2010, MFDN:CPE:2014}, a hybrid MPI/OpenMP code for \textit{ab initio} nuclear structure calculations, using the Daejeon16 NN interaction~\cite{SHIROKOV:2016} and HO basis spaces up through the cutoff $N_{\rm max} = 18$. The dimension of the resulting many-body Hamiltonian matrix is about $2.8 \times 10^9$ at this cutoff.  We note that NCSM basis spaces for $^6\mathrm{Li}$ have now been achieved up through $N_{\rm max} = 22$ in~\cite{Forssen:2017wei}.

This research extends the work presented in~\cite{Negoita:ANN:Li6:2018} where we initially considered the GS energy and GS point-proton rms radius for $^6\mathrm{Li}$ produced with the feed-forward ANN method.
In particular, the current work presents results using multiple datasets, which consist of data through a succession of cutoffs: $N_{\rm max} = 10, 12, 14, 16$, and 18. The previous work considered only one dataset up through $N_{\rm max}$ = 10.  Furthermore, the current  work is the first to report uncertainty assessments of the results. Comparisons of the ANN results and their uncertainties with other extrapolation methods are also provided.

The paper is organized as follows. In Section~\ref{sec:TheoreticalFramework}, short introductions to the \textit{ab initio} NCSM method and ANN's formalism are given. In Section~\ref{sec:ANNDesignFiltering}, our ANN's architecture and filtering are presented. Section~\ref{sec:Results} presents the results and discussions of this work. Section~\ref{sec:Conclusion} contains our conclusion and future work.

\section{Theoretical Framework}\label{sec:TheoreticalFramework}
The NCSM is an \textit{ab initio} approach to the nuclear many-body problem, which solves for the properties of nuclei for an arbitrary inter-nucleon interaction, preserving all the symmetries. The inter-nucleon interaction can consist of both NN components and three-nucleon forces but we omit the latter in the current effort since they are not expected to be essential to the main thrust of the current ANN application.
We will show that the ANN method is useful to make predictions for the GS energy and the GS point-proton rms radius and their extrapolation uncertainties at ultra-large basis spaces using available data from NCSM calculations at smaller basis spaces. More discussions on the NCSM and the ANN are presented in each subsection.
\subsection{\textit{Ab Initio} NCSM Method}\label{sec:NCSM}
In the NCSM method, a nucleus consisting of $A$ nucleons with $N$ neutrons and $Z$ protons ($A = N+Z$)  is described by the quantum Hamiltonian with kinetic energy ($T_{\rm rel}$) and interaction ($V$) terms
\begin{equation}\label{eq:Hamiltonian}
\begin{split}
	& H_A = T_{\rm rel} + V \\ 
	& = \frac{1}{A} \sum_{i<j} \frac{(\vec{p}_i - \vec{p}_j)^2}{2m} + \sum_{i<j}^{A}V_{ij} + \sum_{i<j<k}^{A}V_{ijk} + ~ \ldots.
\end{split}
\end{equation} 
Here, $m$ is the nucleon mass (taken as the average of the neutron and proton mass),  $\vec{p}_i$ is the momentum of the $i$th nucleon, $V_{ij}$ is the NN interaction including the Coulomb interaction between protons, $V_{ijk}$ is the three-nucleon interaction, and the interaction sums run over all pairs and triplets of nucleons, respectively. Higher-body (up to $A$-body) interactions are also allowed and signified by the three dots. As mentioned, we retain only the NN interaction for which we select the Daejeon16 interaction~\cite{SHIROKOV:2016} in the present work.

Our chosen NN interaction, Daejeon16~\cite{SHIROKOV:2016}, is developed from an initial Chiral NN interaction at the next-to-next-to-next-to-leading order (N3LO)~\cite{ENTEM200293, PhysRevC.68.041001} by a process of similarity renormalization group evolution (SRG)~\cite{Bogner:2006pc, Bogner:2009bt} and phase-equivalent transformations (PETs)~\cite{Lurie1997, Lurie2008, PhysRevC.70.044005}.  The PETs are chosen so that Daejeon16 describes well the properties of light nuclei without explicit use of three-nucleon or higher-body interactions, which, if retained, would require a significant increase of computational resources.

With the nuclear Hamiltonian~(\ref{eq:Hamiltonian}), the NCSM solves the $A$-body Schr\"{o}dinger equation
\begin{equation}\label{eq:Schrodinger}
	H_A \Psi_A(\vec{r}_1, \vec{r}_2, \ldots, \vec{r}_A) = E \Psi_A(\vec{r}_1, \vec{r}_2, \ldots, \vec{r}_A), 
\end{equation}
using a matrix formulation, where the $A$-body wave function is given by a linear combination of Slater determinants $\Phi_k$
\begin{equation}\label{eq:Wavefunction}
	\Psi_A(\vec{r}_1, \vec{r}_2, \ldots, \vec{r}_A) = \sum_{k=0}^{n_b} c_k \Phi_k(\vec{r}_1, \vec{r}_2, \ldots, \vec{r}_A),
\end{equation}
and where $n_b$ is the number of many-body basis states, configurations, in the system. 
The Slater determinant $\Phi_k$ is the antisymmetrized product of single-particle wave functions
\begin{equation}\label{eq:SlaterDeterminant}
	\Phi_k(\vec{r}_1, \vec{r}_2, \ldots, \vec{r}_A) = \mathcal{A} \left[\prod_{i=1}^A \phi_{n_il_ij_im_i}(\vec{r}_i)\right],
\end{equation}
where $\phi_{n_il_ij_im_i}(\vec{r}_i)$ is the single-particle wave function for the $i$th nucleon and $\mathcal{A}$ is the antisymmetrization operator. 
Although we adopt a common choice for the single-particle wave functions, the HO basis functions, one can extend this approach to a more general single-particle basis~\cite{NEGOITA:2010, PhysRevC.86.034312, PhysRevC.Caprio:2014iha, Constantinou:2016urz}. The single-particle wave functions are labeled by the quantum numbers $n_il_ij_im_i$, where $n_i$ and $l_i$ are the radial and orbital HO quantum numbers (with $N_i = 2n_i + l_i$ the number of HO quanta for a single-particle state), $j_i$ is the total single-particle angular momentum, and $m_i$ its projection along the $z$ axis.

We employ the $m$ scheme where each HO single-particle state has its orbital and spin angular momenta coupled to good total angular momentum, $j_i$, and magnetic projection, $m_i$.
The many-body basis states $\Phi_k$ have well-defined parity and total angular momentum projection, ${\displaystyle M = \sum_{i=1}^A m_i}$, but they do not have a well-defined total angular momentum $J$.
The matrix elements of the Hamiltonian in the many-body HO basis are given by $H_{ij} = \langle \Phi_i|\hat{H}|\Phi_j \rangle$. These Hamiltonian matrices are sparse, the number of non-vanishing matrix elements follows an approximate scaling rule of $D^{3/2}$, where $D$ is the dimension of the matrix~\cite{Vary:2009qp}. For these large and sparse Hamiltonian matrices, the Lanczos method is one possible choice to find the extreme eigenvalues~\cite{PARLETT:SymmetricEigenvalueProblem:1998}.

We adopt the Lipkin-Lawson method~\cite{PhysRev.109.2071, SpuriousCMMotion:1974} to enforce the factorization of the center-of-mass (CM) and intrinsic components of the many-body eigenstates. In this method, a Lagrange multiplier term, $\lambda(H_{CM} - \frac {3}{2}\hbar\Omega)$, is added to the Hamiltonian above, where $H_{CM}$ is the HO Hamiltonian for the CM motion. With $\lambda$ chosen positive (10 is a typical value), one separates the states of lowest CM motion from the states with excited CM motion by a scale factor of order $\lambda\hbar\Omega$. 

In our $N_{\rm max}$ truncation approach, all possible configurations with $N_{\rm max}$ excitations above the unperturbed GS (the HO configuration with the minimum HO energy defined to be the $N_{\rm max} = 0$ configuration) are considered. The basis is limited to many-body basis states with total many-body HO quanta, ${\displaystyle N_{\rm tot} = \sum_{i=1}^A N_i \leq N_0 + N_{\rm max}}$, where $N_0$ is the minimal number of quanta for that nucleus, which is 2 for $^6\mathrm{Li}$.
Note that this truncation, along with the Lipkin-Lawson approach described above, leads to an exact factorization of the single-particle wave functions into the CM and intrinsic components.
Usually, the basis includes either only many-body states with even values of $N_{\rm tot}$ (and, respectively, $N_{\rm max}$), which correspond to states with the same (positive for $^6\mathrm{Li}$) parity as the unperturbed GS, and are called the ``natural" parity states, or only with odd values of $N_{\rm tot}$ (and, respectively, $N_{\rm max}$), which correspond to states with ``unnatural" (negative for $^6\mathrm{Li}$) parity.

As it was already mentioned, the NCSM calculations are performed with the code MFDn~\cite{MFDN:STERNBERG:2008, MFDN:MARIS:2010, MFDN:CPE:2014}. 
Due to the strong short-range correlations of nucleons in a nucleus, a large basis space is required to achieve convergence. The requirement to simulate the exponential tail of a quantum bound state with HO wave functions possessing Gaussian tails places additional demands on the size of the basis space. The calculations that achieve the desired convergence are often not feasible due to the nearly exponential growth in matrix dimension with increasing $N_{\rm max}$.
To obtain the GS energy and other observables as close as possible to the exact results one seeks solutions in the largest feasible basis spaces. 
These results are sometimes used in attempts to extrapolate to the infinite basis space. To take the infinite matrix limit, several extrapolation methods have been developed, such as ``Extrapolation B"~\cite{ExtrapolationB1:2009, ExtrapolationB2:2013}, ``Extrapolation A5", ``Extrapolation A3", and ``Extrapolation based on $\mathrm{L}_\mathrm{eff}$"~\cite{ExtrapolationA:2017}, which are extensions of techniques developed in~\cite{PhysRevC.86.054002, PhysRevC.86.031301, PhysRevC.87.044326, PhysRevC.91.061301}.  We also note that theoretical extrapolation methods have been introduced and analyzed for quadrupole moments and $E2$ transitions in~\cite{Odell:2015xlw} and for capture cross sections in~\cite{Acharya:2016rek}.   Using such extrapolation methods, one investigates the convergence pattern with increasing basis space dimensions and thus obtains, to within quantifiable uncertainties, results corresponding to the complete basis. We will employ these extrapolation methods to compare with results from ANNs.

\subsection{Artificial Neural Networks}\label{sec:ANNs} 
ANNs are powerful tools that can be used for function approximation, classification, and pattern recognition, such as finding clusters or regularities in the data. The goal of ANNs is to find a solution efficiently when algorithmic methods are computationally intensive or do not exist. 
An important advantage of ANNs is the ability to detect complex non-linear input-output relationships. For this reason, ANNs can be viewed as universal non-linear function approximators~\cite{HORNIK:1989}. Employing ANNs for mapping complex non-linear input-output problems offers a significant advantage over conventional techniques, such as regression techniques, because ANNs do not require explicit mathematical functions.

ANNs are computer algorithms inspired by the structure and function of the brain. Similar to the human brain, ANNs can perform complex tasks, such as learning, memorizing, and generalizing. They are capable of learning from experience, storing knowledge, and then applying this knowledge to make predictions. 

ANNs consist of a number of highly interconnected artificial neurons (ANs), which are processing units. The ANs are connected with each other via adaptive synaptic weights. The AN collects all the input signals and calculates a net signal as the weighted sum of all input signals. Next, the AN calculates and transmits an output signal, $y$. The output signal is calculated using a function called an activation or transfer function, $f$, which depends on the value of the net signal, $y = f(\mathrm{net})$. 

One simple way to organize ANs is in layers, which gives a class of ANN called multi-layer ANN. ANNs are composed of an input layer, one or more hidden layers, and an output layer. The neurons in the input layer receive the data from outside and transmit the data via weighted connections to the neurons in the first hidden layer, which, in turn, transmit the data to the next layer. Each layer transmits the data to the next layer. Finally, the neurons in the output layer give the results. The type of ANN, which propagates the input through all the layers and has no feed-back loops is called a feed-forward multi-layer ANN. For simplicity, throughout this paper we adopt and work with a feed-forward ANN. For other types of ANN, see~\cite{BISHOP:ANN:1995, HAYKIN:ANN:1999}. 

For function approximation, a sigmoid or sigmoidlike and linear activation functions are usually used for the neurons in the hidden and output layer, respectively. There is no activation function for the input layer. The neurons with nonlinear activation functions allow the ANN to learn nonlinear and linear relationships between input and output vectors. Therefore, sufficient neurons should be used in the hidden layer in order to get a good function approximation. 

In our terminology, an ANN is defined by its architecture, the specific values for its weights and biases, and by the chosen activation function. For the purposes of our statistical analysis, we create an ensemble of ANNs.

The development of an ANN is a two-step process with training and testing stages. In the training stage, the ANN adjusts its weights until an acceptable error level between desired and predicted outputs is obtained. The difference between desired and predicted outputs is measured by the error function, also called the performance function. A common choice for the error function is mean-square-error (MSE), which we adopt here.

There are multiple training algorithms based on various implementations of the back-propagation algorithm~\cite{HAGAN:backpropagation:1994}, an efficient method for computing the gradient of error functions. These algorithms compute the net signals and outputs of each neuron in the network every time the weights are adjusted, the operation being called the forward pass operation. Next, in the backward pass operation, the errors for each neuron in the network are computed and the weights of the network are updated as a function of the errors until the stopping criterion is satisfied. 
In the testing stage, the trained ANN is tested over new data that were not used in the training process.

One of the known problems for ANN is overfitting: the error on the training set is within the acceptable limits, but when new data is presented to the network the error is large. In this case, ANN has memorized the training examples, but it has not learned to generalize to new data. This problem can be prevented using several techniques, such as early stopping and different regularization techniques~\cite{BISHOP:ANN:1995, HAYKIN:ANN:1999}.

Early stopping is widely used. In this technique the available data is divided into three subsets: the training set, the validation set, and the test set. The training set is used for computing the gradient and updating the network weights and biases. The error on the validation set is monitored during the training process. When the validation error increases for a specified number of iterations, the training is stopped, and the weights and biases at the minimum of the validation error are returned. The test set error is not used during training, but it is used as a further check that the network generalizes well and to compare different ANN models.

Regularization modifies the performance function by adding a term that consists of the mean of the sum of squares of the network weights and biases. However, the problem with regularization is that it is difficult to determine the optimum value for the performance ratio parameter. It is desirable to determine the optimal regularization parameters automatically. One approach to this process is the Bayesian regularization of MacKay~\cite{MacKay:bayesianinterpolation:1992} that we adopt here as an improvement on early stopping. The Bayesian regularization algorithm updates the weight and bias values according to Levenberg-Marquardt~\cite{HAGAN:backpropagation:1994, Marquardt:LM:1963} optimization. It minimizes a linear combination of squared errors and weights and it also modifies the regularization parameters of the linear combination to generate a network that generalizes well. See~\cite{MacKay:bayesianinterpolation:1992, FORESEE:1997} for more detailed discussions of Bayesian regularization. 
For further and general background on the ANN and how to prevent overfitting and improve generalization refer to~\cite{BISHOP:ANN:1995, HAYKIN:ANN:1999}.

\section{ANN Design and Filtering} \label{sec:ANNDesignFiltering}
\begin{figure}[htbp]
\includegraphics[width=\linewidth]{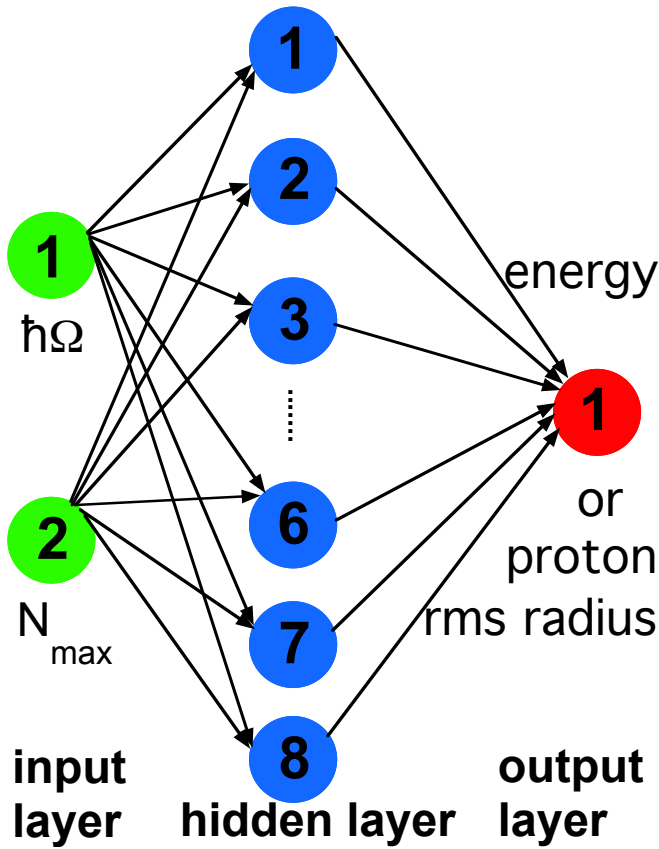} 
\caption{Topological structure of the designed ANN.}\label{fig:ANN_architecture}
\end{figure}

The topological structure of ANNs used in this study is presented in Figure~\ref{fig:ANN_architecture}. The designed ANNs contain one input layer with two neurons, one hidden layer with eight neurons and one output layer with one neuron. The inputs were the basis space parameters: the HO energy, $\hbar\Omega$, and the basis truncation parameter, $N_{\rm max}$, described in Section~\ref{sec:NCSM}. The desired outputs were the GS energy and the GS point-proton rms radius. Separate ANNs were designed for each output. The optimum number of neurons in the hidden layer was obtained according to a trial and error process. The activation function employed for the hidden layer was a widely-used form, the hyperbolic tangent sigmoid function
\begin{equation}\label{eq:tansig}
	f(x) = \text{tansig}(x) = \frac{2}{(1 + e^{-2x})} - 1. 
\end{equation}
It has been proven that one hidden layer and sigmoidlike activation function in this layer are sufficient to approximate any continuous real function, given sufficient number of neurons in the hidden layer~\cite{CYBENKO:1989}.

Every ANN creation and initialization function starts with different initial conditions, such as initial weights and biases and different division of the training, validation, and test datasets. These different initial conditions can lead to very different solutions for the same problem. Moreover, it is also possible to fail to obtain realistic solutions with ANNs for certain initial conditions. For this reason, it is a good idea to train many networks and choose the networks with best performance function values to make further predictions. The performance function, the MSE in our case, measures how well ANN can predict data, i.e., how well ANN can be generalized to new data. 
The test datasets are a good measure of generalization for ANNs since they are not used in training. A small value on the performance function on the test dataset indicates an ANN with good performance was found. However, every time the training function is called, the network gets a different division of the training, validation, and test datasets. That is why, the test sets selected by the training function are a good measure of predictive capabilities for each respective network, but not for all the networks. 

MATLAB software v9.4.0 (R2018a) with NEURAL NETWORK TOOLBOX was used for the implementation of this work. As mentioned before in Section~\ref{sec:Introduction}, the application here is the $^6\mathrm{Li}$ nucleus. The dataset was generated with the \textit{ab initio} NCSM calculations using the MFDn code with the Daejeon16 NN interaction~\cite{SHIROKOV:2016} and a sequence of basis spaces up through $N_{\rm max} = 18$. The $N_{\rm max} = 18$ basis space corresponds to our largest matrix diagonalized using the \textit{ab initio} NCSM approach for $^6\mathrm{Li}$ with dimension of about $2.8 \times 10^9$. Only the ``natural" parity states, which have even $N_{\rm max}$ values for $^6\mathrm{Li}$, were considered in this work.

For our application here, we choose to compare the performance for all the networks by taking the original dataset and dividing it into a design set and a test set. The design (test) set consists of 16/19 (3/19) of the original dataset. The design set is further randomly divided by the \textit{train} function into a training set and another test set. This training (test) set comprises 90\% (10\%) of the design set. 

For each design set, we train 100 ANNs with the above architecture and with each ANN starting from different initial weights and biases. To ensure good generalization, each ANN is retrained ten times, during which we sequentially evolve the weights and biases. A back-propagation algorithm with Bayesian regularization with MSE performance function was used for ANN training. Bayesian regularization does not require a validation dataset. 

For function approximation, Bayesian regularization provides better generalization performance than early stopping in most cases, but it takes longer to converge to the desired performance ratio. The performance improvement is more noticeable when the dataset is small because Bayesian regularization does not require a validation dataset, leaving more data for training. 
In MATLAB, Bayesian regularization has been implemented in the function \textit{trainbr}. When using \textit{trainbr}, it is important to train the network until it reaches convergence.
In this study, the training process is stopped if: (i) it reaches the maximum number of iterations, 1000; (ii) the performance has an acceptable level; (iii) the estimation error is below the target; or (iv) the Levenberg-Marquardt adjustment parameter $\mu$ becomes larger than $10^{10}$. A typical indication for convergence is when the maximum value of $\mu$ has been reached.

In order to develop confidence in our ANNs, we organize a sequence of challenges consisting of choosing original datasets that have successively improved information originating from NCSM calculations. That is, we define an ``original dataset" to consist of NCSM results at 19 selected values of $\hbar\Omega = 8, 9, 10$ $MeV$ and then in 2.5 $MeV$ increments covering 10 -- 50 $MeV$ for all $N_{\rm max}$ values up through, for example, 10 (our first original dataset). We define our second original dataset to consist of NCSM results at the same values of $\hbar\Omega$ but for all $N_{\rm max}$ values up through 12. We continue to define additional original datasets until we have exhausted available NCSM results at $N_{\rm max} = 18$.

To split each original dataset (defined by its cutoff $N_{\rm max}$ value) into 16/19 and 3/19 subsets we randomly choose three points for each $N_{\rm max}$ value within the cutoff $N_{\rm max}$ value.
The resulting 3/19 set is our test set used to subselect optimum networks from these 100 ANNs. Figure~\ref{alg:networksFiltering} shows the general procedure for selecting the ANNs used to make predictions for nuclear physics observables, where ``test1" is the 3/19 test set described above. We retain only those networks which have a MSE on the 3/19 test set below 0.002 $MeV^2$ ($5.0 \times 10^{-6}~fm^2$) for the GS energy (GS point-proton rms radius). 
We then cycle through this entire procedure with a specific original dataset 400 times in order to obtain an estimated 50 ANNs that would satisfy additional screening criteria. That is, the retained networks are further filtered based on the following criteria.
\begin{enumerate}[(i)]
	\item The networks must have a MSE on their design set below 0.0002 $MeV^2$ ($5.0 \times 10^{-7}~fm^2$) for the GS energy (GS point-proton rms radius).
\begin{figure*}[!htb]
\begin{lstlisting}[mathescape=true] % caption=General procedure for selecting ANNs used to make predictions for nuclear physics observables., captionpos=b, label={alg:networksFiltering}

for each observable do
   for each original dataset do
      repeat
         for trial=1:400 do
            initialize test1
            initialize design = original\test1
            for each network of 100 networks do
               initialize network
      	       for i=1:10 do
                  train network
                  if i == 1 then
                     smallest = MSE(test1) 
                        if MSE(test1) > val$_1$ then
                           break
                        end if
                  else
                     if MSE(test1) < smallest
                        smallest = MSE(test1)
                     end if
                  end if
               end for
               if i $\neq$ 1 then
                  save network with MSE(test1) = smallest into saved_networks1
               end if
            end for
         end for
         % networks further filtering
         for each network in saved_networks1 do
            if MSE(design) $\leq$ val$_2$ then
               save network in saved_networks2
               if observable == GS energy then
                  check variational principle
                  if !(variational principle) then
                     remove network from saved_networks2
                  end if
               end if
            end if
         end for
         sort saved_networks2 based on MSE(design)
         numel = min(50, length(saved_networks2))
         networks_to_predict = saved_networks2(1:numel)
         % discard elements lying outside three-sigma of their mean
         apply three-$\sigma$ rule to networks_to_predict
         if numel == 50 and length(networks_to_predict) $<$ 50 then
            repeat           
               add next element from saved_networks2 to networks_to_predict            
               apply three-$\sigma$ rule to networks_to_predict
            until !exist elements in saved_networks2 or length(networks_to_predict) == 50
         end if            
      until length(networks_to_predict) == 50
   end for
end for
\end{lstlisting}
\caption{General procedure for selecting ANNs used to make predictions for nuclear physics observables.}
\label{alg:networksFiltering}
\end{figure*}
	\item For the GS energy, the networks' predictions should satisfy the theoretical physics upper-bound (variational) condition for all increments in $N_{\rm max}$ up to $N_{\rm max} = 70$. That is the ANNs predictions for the GS energy should decrease uniformly with increasing $N_{\rm max}$ up to $N_{\rm max} = 70$. All ANNs at this stage of filtering were found to satisfy this criteria so no ANNs were rejected according to this condition.
	\item Pick the best 50 networks based on their performance on the design set which satisfy a three-$\sigma$ rule: the predictions at $N_{\rm max} = 70$ ($N_{\rm max} = 90$) for the GS energy (GS point-proton rms radius) produced by these 50 networks are required to lie within three standard deviations (three-$\sigma$) of their mean.  Thus, predictions lying outside three-$\sigma$ are discarded as outliers. This is an iterative method since a revised standard deviation could lead to the identification of additional outliers.  The three-$\sigma$ method was initially proposed in~\cite{PhysRevC.Gross:2008ps} and then implemented by the Granada group for analysis of NN scattering data~\cite{PhysRevC.Perez:2013jpa}.
\end{enumerate}
If, at this stage, we obtained less than 50 networks in our statistical sample we go through the entire procedure with that specific original dataset an additional 400 times. In no case did we find it necessary to run more than 1200 cycles. 

\section{Results and Discussions} \label{sec:Results}
This section presents $^6\mathrm{Li}$ results along with their estimated uncertainties for the GS energy and point-proton rms radius using the feed-forward ANN method. Comparison with results from other extrapolation methods is also provided.
Preliminary results of this study were presented in~\cite{Negoita:ANN:Li6:2018}. The results of this work extend the preliminary results as follows: multiple original datasets up through a succession of cutoffs: $N_{\rm max} = 10, 12, 14, 16$, and 18 are used to design, train and test the networks; for each original dataset, 50 best networks are selected using the methodology described in Section~\ref{sec:ANNDesignFiltering} and the distribution of the results is presented as input for the uncertainty assessment.

The 50 selected ANNs for each original dataset were used to predict the GS energy at $N_{\rm max} = 70$ and the GS point-proton rms radius at $N_{\rm max} = 90$ for 19 aforementioned values of $\hbar\Omega = 8-50$ $MeV.$  These ANN predictions were found to be approximately independent of $\hbar\Omega$.
The ANN estimate of the converged result, i.e., the result from an infinite matrix, was taken to be the median of the predicted results at $N_{\rm max} = 70$ ($N_{\rm max} = 90$) over the 19 selected values of $\hbar\Omega$ for each original dataset. 

\begin{figure*}[!htbp]
\centering%
\includegraphics[width=\textwidth]{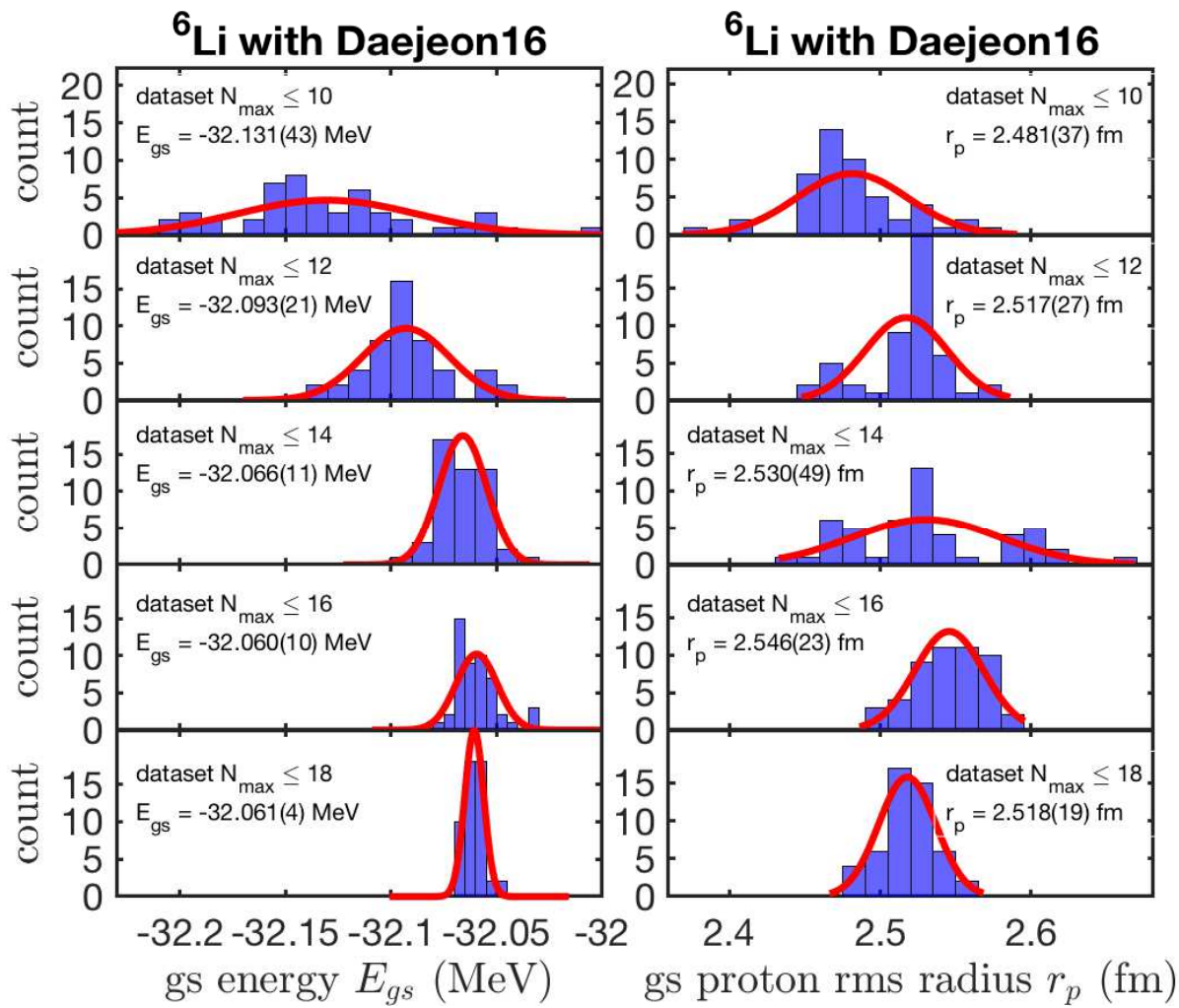} 
\caption{Statistical distributions of the predicted GS energy (left) and GS point-proton rms radius (right) of $^6\mathrm{Li}$ produced by ANNs trained with NCSM simulation data at increasing levels of truncation up to $N_{\rm max} = 18$. The ANN predicted GS energy (GS point-proton rms radius) is obtained at  $N_{\rm max} = 70~(90)$. The extrapolates are quoted for each plot along with the uncertainty indicated in parenthesis as the amount of uncertainty in the least significant figures quoted.}\label{fig:Li6_histogram}
\end{figure*}

In order to obtain the uncertainty assessments of the results, we constructed a histogram with a normal (Gaussian) distribution fit to the results predicted by the 50 selected ANNs for each original dataset and for each observable. Figure~\ref{fig:Li6_histogram} presents these histograms along with their corresponding Gaussian fits. The cutoff value of $N_{\rm max}$ in each original dataset used to design, train and test the networks is indicated on each plot along with the parameters used in fitting: the mean ($\mu = E_{\rm GS}$ or $r_{\rm p}$) and the quantified uncertainty ($\sigma$) indicated in parenthesis as the amount of uncertainty in the least significant figures quoted. The mean values ($\mu = E_{\rm GS}$ or $r_{\rm p}$) represent the extrapolates obtained using the feed-forward ANN method. It is evident from the Gaussian fits in Figure~\ref{fig:Li6_histogram} that, as we successively expand the original dataset to include more information originating from NCSM calculations by increasing the cutoff value of $N_{\rm max}$ in the dataset,  the uncertainty generally decreases. Furthermore, there is apparent consistency with increasing cutoff $N_{\rm max}$ since successive extrapolates are consistent with previous extrapolates within the assigned uncertainties for each observable. An exception is the GS point-proton rms radius when using the original dataset with cutoff $N_{\rm max}=14$. In this case, note the single Gaussian distribution exhibits an uncertainly much bigger than the case with cutoff $N_{\rm max}=12$. 
The histogram for $r_p$ at cutoff $N_{\rm max}=14$ shows a hint of multiple peaks, which could indicate multiple local minima within the limited sample of 50 ANNs.

Upon further inspection of Figure~\ref{fig:Li6_histogram}, one may question whether a Gaussian approximately
represents all sets of histograms.  Let us consider the five cases, which stand out in
this regard: the GS energy for $N_{\rm max} \leq 10$
and $N_{\rm max} \leq 16$ as well as the point-proton rms radius for $N_{\rm max} \leq 10$, $N_{\rm max} \leq 12$,
and $N_{\rm max} \leq 14$. These five cases exhibit gaps and outliers more prominently than
the remaining cases. For these five cited cases, with the exception
of the $N_{\rm max} \leq 14$ point-proton rms radius case discussed above, we do find  
that at least 63\% of the ANNs lie within their quoted 1-$\sigma$ value and at least 86\% 
lie within their quoted 2-$\sigma$ value. Owing to the method of discarding outliers 
described above, all 50 ANNs fall within their quoted 3-$\sigma$ values. These 
characteristics of the distributions of our ANN results lend support to our 
Gaussian fit procedure.

It is worth noting that the widths of the Gaussian fits to the histograms suggest that there is a larger relative uncertainty of the point-proton radius extrapolation than that of the GS energy extrapolation produced by the ANNs.  In other words, as one proceeds down the five panels in Figure~\ref{fig:Li6_histogram} from the top, the uncertainty in the GS energy decreases significantly faster than the uncertainty in the point-proton radius.  This reflects the well-known feature of NCSM results in a HO basis where long-range observables, such as $r_p$, are more sensitive than the GS energy to the slowly converging asymptotic tails of the nuclear wave function.

\begin{figure}[!htbp]
\includegraphics[width=\linewidth]{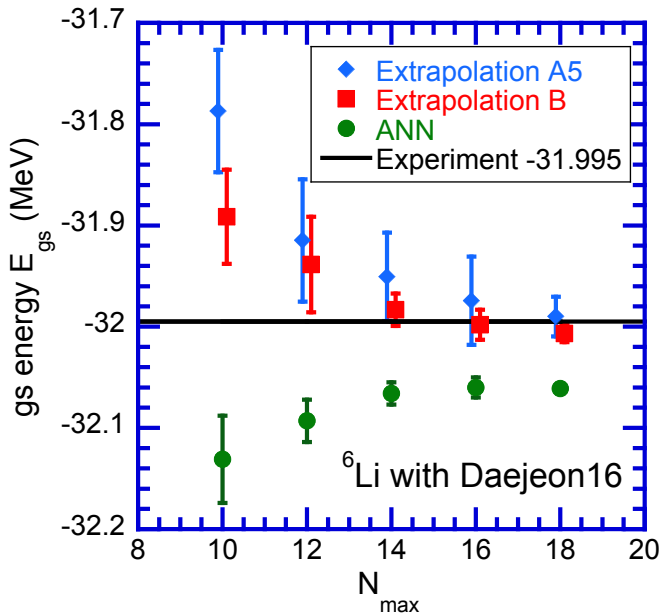} 
\caption{(Color online) Extrapolated GS energies of $^6\mathrm{Li}$ with Daejeon16 using the feed-forward ANN method (green), the ``Extrapolation A5"~\cite{ExtrapolationA:2017} method (blue) and the ``Extrapolation B"~\cite{ExtrapolationB1:2009, ExtrapolationB2:2013} method (red) as a function of the cutoff value of $N_{\rm max}$ in each dataset. Error bars represent the uncertainties in the extrapolations. The experimental result is also shown by the black horizontal solid line~\cite{TILLEY20023}.}\label{fig:Li6_gs_energy_extrap}
\end{figure}

Figure~\ref{fig:Li6_gs_energy_extrap} presents the sequence of extrapolated results for the GS energy using the feed-forward ANN method in comparison with results from ``Extrapolation A5"~\cite{ExtrapolationA:2017} and ``Extrapolation B"~\cite{ExtrapolationB1:2009, ExtrapolationB2:2013} methods. Uncertainties are indicated as error bars and are quantified using the rules from the respective procedures. The experimental result is also shown by the black horizontal solid line~\cite{TILLEY20023}. The ``Extrapolation B" method adopts a three-parameter extrapolation function that contains a term that is exponential in $N_{\rm max}$. The ``Extrapolation A5" method adopts a five-parameter extrapolation function that contains a term that is exponential in $\sqrt{N_{\rm max}}$ in addition to the single exponential in $N_{\rm max}$ used in the ``Extrapolation B" method. Note in Figure~\ref{fig:Li6_gs_energy_extrap} the convergence pattern for the GS energy with increasing cutoff $N_{\rm max}$ values. All extrapolation methods provide their respective error bars which generally decrease with increasing cutoff $N_{\rm max}$.
Also note the visible upward trend for the extrapolated energies when using the feed-forward ANN method while there is a downward trend for the ``Extrapolation A5" and ``Extrapolation B" methods. While these smooth trends in the extrapolated results of Figure~\ref{fig:Li6_gs_energy_extrap} suggest systematic errors are present in each method, the quoted uncertainties are large enough to nearly cover the systematic trends displayed within each method but the quoted uncertainties are not sufficient to cover the differences between the methods.

\begin{figure}[!htbp]
\includegraphics[width=\linewidth]{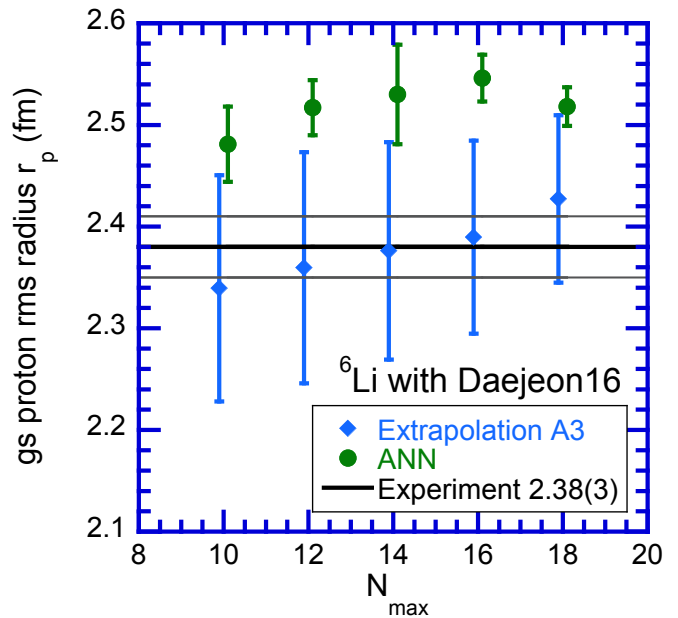} 
\caption{(Color online) Extrapolated GS point-proton rms radii of $^6\mathrm{Li}$ with Daejeon16 using the feed-forward ANN method (green) and the ``Extrapolation A3"~\cite{ExtrapolationA:2017} method (blue) as a function of the cutoff value of $N_{\rm max}$ in each dataset. Error bars represent the uncertainties in the extrapolations. The experimental result and its uncertainty are also shown by the horizontal lines~\cite{TANIHATA2013215}.}
\label{fig:Li6_gs_proton_radius_extrap}
\end{figure}

Figure~\ref{fig:Li6_gs_proton_radius_extrap} presents the sequence of extrapolated results for the GS point-proton rms radius using the feed-forward ANN method in comparison with results from ``Extrapolation A3"~\cite{ExtrapolationA:2017} method. The ``Extrapolation A3" method adopts a different three-parameter extrapolation function than the ``Extrapolation A5" method used for the GS energy. For the GS point-proton rms radius there is mainly a systematic upward trend in the extrapolations and the uncertainties are only decreasing slowly with cutoff $N_{\rm max}$ when using the ``Extrapolation A3" method. However, when using the feed-forward ANN method, the predicted rms radius increases until cutoff $N_{\rm max} = 16$ and then decreases again. The experimental result is shown by the bold black horizontal line and its error band is shown by the thin black lines above and below the experimental line. We quote the experimental value for the GS point-proton rms radius that has been extracted from the measured charge radius by applying established electromagnetic corrections~\cite{TANIHATA2013215}.

While the extrapolation results from the ANNs show reasonable consistency with each other as a function of increasing the cutoff $N_{\rm max}$ of the training data sets, there are trends in these extrapolations suggesting that systematic uncertainties are also present in the ANN predictions.  Note that the analytical functions employed for extrapolations show trends suggesting that they also have systematic uncertainties.  As a consequence, one can surmise that results presented in Figures~\ref{fig:Li6_gs_energy_extrap} and~\ref{fig:Li6_gs_proton_radius_extrap} suggest that all results would be more consistent with each other if their current internal estimates of uncertainties were at least doubled to encompass the role of their respective, but as yet unquantified, systematic uncertainties. However, our comparisons in Figures~\ref{fig:Li6_gs_energy_extrap} and~\ref{fig:Li6_gs_proton_radius_extrap} are not sufficient to indicate a quantitative systematic uncertainty for each method.   Rather, we employ the present comparisons to reveal the likely presence of systematic uncertainties in the compared methods and suggest a comprehensive study of results from multiple nuclei and different interactions will be needed to fully quantify the systematic uncertainties of each method.

The extrapolated results along with their uncertainty estimations for the GS energy and the GS point-proton rms radius of $^6\mathrm{Li}$ and the variational upper bounds for the GS energy are also quoted in Table~\ref{tab:results}. The extrapolation arises when using all available results up through the cutoff $N_{\rm max}$ values shown in the table. All the extrapolated energies were below their respective variational upper bounds. Our current results, taking into consideration our assessed uncertainties, appear to be reasonably consistent with the results of the single ANN using the dataset up through the cutoff $N_{\rm max} = 10$ developed in~\cite{Negoita:ANN:Li6:2018}. Also note the feed-forward ANN method produces smaller uncertainty estimations than the other extrapolation methods.  In addition, as seen in Figures~\ref{fig:Li6_gs_energy_extrap} and~\ref{fig:Li6_gs_proton_radius_extrap}, the ANN predictions imply that Daejeon16 provides converged results slightly further from experiment than the other extrapolation methods.

\begin{table*}[!htb] 
  \caption{
Comparison of the ANN predicted results with results from the current best upper bounds and from other extrapolation methods, such as Extrapolation A$^\mathrm{a}$ ~\cite{ExtrapolationA:2017} and Extrapolation B~\cite{ExtrapolationB1:2009, ExtrapolationB2:2013}, with their uncertainties. The experimental GS energy is taken from~\cite{TILLEY20023}. The experimental point-proton rms radius is obtained from the measured charge radius by the application of electromagnetic corrections~\cite{TANIHATA2013215}. Energies are given in units of $MeV$ and radii are in units of femtometers (\textit{fm}).} 
  \label{tab:results}
  \begin{threeparttable}[b]
    \resizebox{\linewidth}{!}{
     \begin{tabular}{c|c|c|c|c|c|c} 
      Observable & Experiment & $N_{\rm max}$ & Upper Bound &  Extrapolation A\tnote{a}~~& Extrapolation B & ANN \\  
      \hline
      GS energy & -31.995 & 10  & -31.688 & -31.787(60) & -31.892(46) & -32.131(43)\\ 
      & & 12  & -31.837 & -31.915(60) & -31.939(47) & -32.093(21)\\
      & & 14  & -31.914 & -31.951(44) & -31.983(16) & -32.066(11)\\
      & & 16  & -31.954 & -31.974(44) & -31.998(15) & -32.060(10)\\
      & & 18  & -31.977 & -31.990(20) & -32.007(9) & -32.061(4)\\
      \hline
      GS point-proton rms radius & 2.38(3) & 10  & -- & 2.339(111) & -- & 2.481(37)\\
       & & 12  & -- & 2.360(114) & -- & 2.517(27)\\
       & & 14  & -- & 2.376(107) & -- & 2.530(49)\\
       & & 16  & -- & 2.390(95) & -- & 2.546(23)\\
       & & 18  & -- & 2.427(82) & -- & 2.518(19)\\
    \end{tabular}
    }
    \begin{tablenotes}[flushleft] 
           \item \makebox[\textwidth] { [a] The ``Extrapolation A5" method for the GS energy and the ``Extrapolation A3" method for the GS point-proton rms radius}
         \end{tablenotes}
  \end{threeparttable}
\end{table*}

\begin{figure}[htbp]
\includegraphics[width=\linewidth]{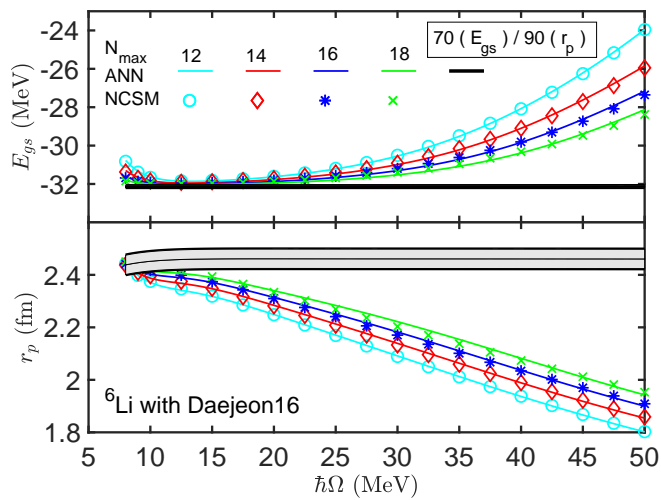} 
\caption{Comparison of the best ANN predictions based on dataset with $N_{\rm max} \leq 10$ and the corresponding NCSM calculated GS energy and GS point-proton rms radius values of $^6\mathrm{Li}$ as a function of $\hbar\Omega$ at $N_{\rm max} = 12, 14, 16,$ and $18$. The shaded area corresponds to the ANN nearly converged result at $N_{\rm max} = 70$ (GS energy) and $N_{\rm max} = 90$ (GS point-proton rms radius) along with its uncertainty estimation quantified as described in the text.
}\label{fig:Li6_observable_cp}
\end{figure}

To illustrate a convergence example, the network with the lowest performance function, i.e., the lowest MSE, using the original dataset at $N_{\rm max} \leq 10$ is selected from among the 50 networks to predict the GS energy (GS point-proton rms radius) for $^6\mathrm{Li}$ at $N_{\rm max} = 12$, 14, 16, 18 and 70 (90). Figure~\ref{fig:Li6_observable_cp} presents these ANN predicted results of the GS energy and point-proton rms radius and the corresponding NCSM calculation results at the available succession of cutoffs: $N_{\rm max} = 12$, 14, 16, and 18 for comparison as a function of $\hbar\Omega$. The solid curves are smooth curves drawn through 100 data points of the ANN predictions and the individual symbols represent the NCSM calculation results. The nearly converged result predicted by the best ANN and its uncertainty estimation, obtained as described in the text above, are also shown by the shaded area at $N_{\rm max} = 70$ and $N_{\rm max} = 90$ for the GS energy and the GS point-proton rms radius, respectively. Figure~\ref{fig:Li6_observable_cp} shows good agreement between the ANN predictions and the calculated NCSM results at $N_{\rm max} = $ 12 -- 18. 

Predictions of the GS energy by the best 50 ANNs converged uniformly with increasing $N_{\rm max}$ down towards the final result.  In addition, these predictions became increasingly independent of the basis space parameters, $\hbar\Omega$ and $N_{\rm max}$.
The ANN is successfully simulating what is expected from the many-body theory applied in a configuration interaction approach. That is, the energy variational principle requires that the GS energy behaves as a non-increasing function of increasing matrix dimensionality at fixed $\hbar\Omega$ (basis space dimension increases with increasing $N_{\rm max}$). That the ANN result for the GS energy is essentially a flat line at $N_{\rm max} = 70$ provides a good indication that the ANN is producing a valuable estimate of the converged GS energy. 

The GS point-proton rms radii provide a dependence on the basis size and $\hbar\Omega$ which is distinctly different from the GS energy in the NCSM.  In particular, these radii are not monotonic with increasing $N_{\rm max}$ at fixed $\hbar\Omega$ and they are more slowly convergent with increasing basis size.
However, the GS point-proton rms radius converges monotonically from below for most of the $\hbar\Omega$ range shown. More importantly, the GS point-proton rms radius also shows the anticipated convergence to a flat line when using the ANN predictions at $N_{\rm max} = 90$.

\section{Conclusion and Future Work}\label{sec:Conclusion}

We used NCSM computational results to train feed-forward ANNs to predict the properties of the $^6\mathrm{Li}$ nucleus,  in particular the converged GS energy and the converged point-proton rms radius along with their quantified uncertainties. The advantage of the ANN method is that it does not need any mathematical relationship between input and output data as opposed to other available extrapolation methods. The architecture of ANNs consisted of three layers: two neurons in the input layer, eight neurons in the hidden layer and one neuron in the output layer. Separate ANNs were designed for each output. 

We have generated theoretical data for $^6\mathrm{Li}$ by performing \textit{ab initio} NCSM calculations with the MFDn code using the Daejeon16 NN interaction and HO basis spaces up through the cutoff $N_{\rm max} = 18$. 

To improve the fidelity of our predictions, we use an ensemble of ANNs obtained from multiple trainings to make predictions for the quantities of interest. This involved developing a sequence of applications using multiple datasets up through a succession of cutoffs. That is, we adopt cutoffs of $N_{\rm max} = 10, 12, 14, 16$, and 18 at 19 selected values of $\hbar\Omega = $ 8 -- 50 $MeV$ to train and test the networks. 

We introduced a method for quantifying uncertainties using the feed-forward ANN method by constructed a histogram with a normal (Gaussian) distribution fit to the converged results predicted by the best performing 50 ANNs. The ANN estimate of the converged result (i.e. the result from an infinite matrix) was taken to be the median of the predicted results at $N_{\rm max} = 70~(90)$ over the 19 selected values of $\hbar\Omega$ for the GS energy (GS point-proton rms radius). The parameters used in fitting the normal distribution were the mean, which represents the extrapolate, and the quantified uncertainty, $\sigma$. 

The designed ANNs were sufficient to produce results for these two very different observables in $^6\mathrm{Li}$ from the \textit{ab initio} NCSM. Through our tests, the ANN predicted results were in agreement with the available \textit{ab initio} NCSM results. The GS energy and the GS point-proton rms radius showed good convergence patterns and satisfied the theoretical physics condition, independence of basis space parameters in the limit of extremely large matrices. 

Comparisons of the ANN results with other extrapolation methods of estimating the results in the infinite matrix limit were also provided along with their quantified uncertainties. The results for ultra-large basis spaces were in approximate agreement with each other. Table~\ref{tab:results} presents a summary of our results, performed with the feed-forward ANN method introduced here, as well as performed with the ``Extrapolations A" and ``Extrapolation B" methods, introduced earlier.

By these measures, ANNs are seen to be successful for predicting the results of ultra-large basis spaces, spaces too large for direct many-body calculations. It is our hope that ANNs will help reap the full benefits of HPC investments. 

As future work, additional $\mathrm{Li}$ isotopes such as $^7\mathrm{Li}$, $^8\mathrm{Li}$, and $^9\mathrm{Li}$, then heavier nuclei, will be investigated using the ANN method and the results will be compared with results from other extrapolation methods. Moreover, this method will be applied to other observables such as magnetic moment, quadruple transition rates, etc.

\section*{Acknowledgment}

This work was supported in part by the Department of Energy under Grants No. DE-FG02-87ER40371 and No. DESC000018223 (SciDAC-4/NUCLEI), and by Professor Glenn R. Luecke's funding at Iowa State University. The work of A.M.S. was supported by the Russian Science Foundation under Project No. 16-12-10048. The work of I.J.S and Y.K. was supported partly by the Rare Isotope Science Project of Institute for Basic Science funded by Ministry of Science, ICT and Future Planning and NRF of Korea (2013M7A1A1075764).  Computational resources were provided by the National Energy Research Scientific Computing Center (NERSC), which is supported by the Office of Science of the U.S. DOE under Contract No. DE-AC02-05CH11231.

\bibliography{bibtemplate_ANN_Nuclear_Physics_2018_short_revtex4}

\end{document}